\documentclass[preprint,showpacs,preprintnumbers,amsmath,amssymb]{revtex4}

\usepackage{graphicx}% Include figure files
\usepackage{dcolumn}% Align table columns on decimal point
\usepackage{bm}% bold math

\def\beq{\begin{eqnarray}}    %%%  begequation/eqnarray
\def\eeq{\end{eqnarray}}      %%%  endequation/eqnarray
\newcommand{\be}{\begin{equation}}
\newcommand{\ee}{\end{equation}}
\newcommand{\ba}{\begin{eqnarray}}
\newcommand{\ea}{\end{eqnarray}}
\newcommand{\bea}{\begin{eqnarray*}}
\newcommand{\eea}{\end{eqnarray*}}

\begin{document}

\title{Constraints on dark energy from the observed density
fluctuations spectrum and supernova data}

\author{Reuven Opher}
\email{opher@astro.iag.usp.br}
\author{Ana Pelinson}
\email{anapel@astro.iag.usp.br} \affiliation{IAG, Universidade de
S\~{a}o Paulo, Rua do Mat\~{a}o, 1226 \\
Cidade Universit\'aria, CEP 05508-900. S\~{a}o Paulo, S.P.,
Brazil}

\date{\today}

\begin{abstract}
%%%%%%%%%%%%%%%%%%%%%%%%%%%%%%%%%%%%%%%%%%%%%%%%%%%%%%%%%%%%%%%
%%%%%%%%%%%%%%%%%%%%%%%%%%%%%%%%%%%%%%%%%%%%%%%%%%%%%%%%%%%%%%%

One of the greatest challenges in cosmology today is to determine the nature of dark
energy, the source of the observed present acceleration of the universe. High
precision experiments are being developed to reduce the uncertainties in the
observations. Recently, we showed that the agreement to an accuracy of $10\%$ of
measurements of the present density fluctuations $(\delta\rho/\rho)^2$, derived from
galaxy distribution (GD) data and cosmic microwave background (CMB) anisotropies in
the $\Lambda$CDM model, puts very strong limits on the possible decay of the vacuum
energy into cold dark matter. Using this agreement, combined with the evidence that
the matter density $\Omega_M^0=0.28\pm 0.02$ and that the universe is approximately
flat, we show that the vacuum metamorphosis model (VMM) and the popular brane-world
model (BWM), both used to explain dark energy, can be discarded. When we relax the
$\Omega_M^0$ requirement, we find that an agreement within $10\%$ can be obtained
only with $\Omega_M^0\simeq 0.36$ for the VMM and $\Omega_M^0\simeq 0.73$ for the
BWM, both of which are not consistent with observations. The agreement of the CMB
and GD data and previous constraints from SNIa data exclude, or put strong limits
on, other dark energy models, which have been suggested, that can be described by
the parametrized equation of state (EOS) $w=p/\rho= w_0 + w_a(1-a)$, where $w_0$ and
$w_a$ are constants, $a$ is the cosmological scale factor and  $p\,(\rho)$ is the
pressure (energy density) of the dark energy. We find that the supergravity (SUGRA)
model with $w_0=-0.82$ and $w_a=0.58$ can be discarded. In general, we find best
values $-1.86<w_0<-1.72$ with $1.53<w_a<2.0$. For redshifts $z\sim 0.5-1$, where the
supernova data is sensitive, $w\sim -1$ for this parametrized EOS.

%%%%%%%%%%%%%%%%%%%%%%%%%%%%%%%%%%%%%%%%%%%%%%%%%%%%%%%%%%%%%%%
%%%%%%%%%%%%%%%%%%%%%%%%%%%%%%%%%%%%%%%%%%%%%%%%%%%%%%%%%%%%%%%
\end{abstract}

\maketitle

%%%%%%%%%%%%%%%%%%%%%%% / * \ %%%%%%%%%%%%%%%%%%%%%%%%
%%%%%%%%%%%%%%%%%%%%%% /*%#%*\ %%%%%%%%%%%%%%%%%%%%%%%
%%%%%%%%%%%%%%%%%%%%%% \*%#%*/ %%%%%%%%%%%%%%%%%%%%%%%
%%%%%%%%%%%%%%%%%%%%%%% \ * / %%%%%%%%%%%%%%%%%%%%%%%%
%%% 1
\section{Introduction}

Many dark energy models have been suggested to explain the recent acceleration of
the universe, first indicated by SNIa observations \cite{Riess,Perl}. The nature of
dark energy is one of the major problems in cosmology. Theories in which gravity is
modified  as well as parametrizations of the dark energy equation of state (EOS)
$w(z)=p/\rho$, where $p\,(\rho)$ is the pressure (energy density) of the dark
energy, have been suggested \cite{wag86,lin88,fpoc}. Based on observations, various
constraints have been put on the EOS for a variety of models (e.g.,
\cite{fhlt,huttur,hutstark,somehu,linsl}).

We begin by analyzing two interesting models that have been suggested for modifying
gravity, which allow for a description of dark energy in terms of an effective EOS.
The first is the five-dimensional brane-world model (BWM) of Defayet et al.
\cite{bwm}, where gravity is modified by adding a five-dimensional Einstein-Hilbert
action that dominates at distances which are greater than the crossover distance
$r_c=M_{Pl}^2/(M_{5}^3)$, where $M_{Pl}$ is the Planck mass and $M_{5}$ is a
five-dimensional Planck mass. The second model is the vacuum metamorphosis model
(VMM) of Parker and Raval \cite{vmm}, which assumes the existence of a quantized
non-interacting scalar field coupled to the Ricci scalar curvature. In the VMM, the
quantum vacuum undergoes a phase transition at a redshift $z_j$\,: from a zero value
at $z>z_j$ to a non-zero value for $z<z_j$.

Linder \cite{lindastro} analyzed the linear growth of a density perturbation
$(\delta\rho/\rho)$ and the gravitational potential for both of these models,
comparing them with the simplest linear parametrization of the dark energy EOS as a
function of $a$\,: $w(a)=w_0+w_a(1-a)$, where $w_0$ and $w_a$ are constants and $a$
is the cosmological scale factor.

Measurements of the present density fluctuations
$(\delta\rho/\rho)^2$, derived from the cosmic microwave background
(CMB) anisotropies in the $\Lambda$CDM model, have been compared
with those derived from the 2dF Galaxy Redshift Survey (2dFGRS)
\cite{2df,2df1}. It was found that their difference $(F)$ is no more
than $10\%$ \cite{2df}. We recently showed that an agreement within
$10\%$ (i.e., $F_{\rm max}=0.1$) of the two sets of
$(\delta\rho/\rho)^2$ puts strong limits on a possible decay of the
vacuum energy into CDM \cite{vdecay}. Here we make a similar
analysis to show that the BWM and the VMM as dark energy candidates
can be discarded, in the face of the present evidence that $F_{\rm
max}=0.1$, the matter density $\Omega_M^0=0.28\pm 0.02$ and that the
universe is flat, indicated by recent CMB data.

We use the agreement to within $10\%$ of the present observed $(\delta\rho/\rho)^2$
between the CMB and galaxy distribution (GD) data and the constraints from the Gold
SNIa data \cite{comp} to restrict the parameters of the EOS, $w(a)=w_0+w_a(1-a)$. In
particular, we analyze a model suggested by supergravity (SUGRA) \cite{sugra},
studied by Linder \cite{lindastro} and recently by Solevi et al. \cite{solevi}).

In $\S$ II, we discuss the effect of dark energy on the linear
growth of $(\delta\rho/\rho)$ in the models: 1)~the vacuum energy
decay into CDM; 2)~the BWM and VMM; 3)~dark energy models
parametrized by $w(a)=w_0+w_a(1-a)$; and 4)~the SUGRA model. Our
conclusions are presented in section III.

%%%%%%%%%%%%%%%%%%%%%%% / * \ %%%%%%%%%%%%%%%%%%%%%%%%
%%%%%%%%%%%%%%%%%%%%%% /*%#%*\ %%%%%%%%%%%%%%%%%%%%%%%
%%%%%%%%%%%%%%%%%%%%%% \*%#%*/ %%%%%%%%%%%%%%%%%%%%%%%
%%%%%%%%%%%%%%%%%%%%%%% \ * / %%%%%%%%%%%%%%%%%%%%%%%%
%%% 2
\section{Dark energy and the growth of density fluctuations}

The nature of dark energy is still unknown and there are many alternative models to
explain it. One possibility is that instead of a constant vacuum energy, described
by a cosmological constant, we have a vacuum energy which is decaying. Other
possibilities are models that modify gravity and phenomenological models that
parametrize the dark energy EOS in the form $w(a)= w_0+w_a (1-a)$, setting values
for $w_0$ and $w_a$. In this section, all of the above models will be analyzed. For
these models, the Friedmann equation can be written in a general form in terms of an
effective EOS \cite{lindmnras}. Modeling the dark energy as an ideal fluid in a flat
universe, we can write the Friedmann equation as
\begin{equation}
\frac{H^2(z)}{H_0^2} = \Omega_{M}^0(1+z)^3 + (1-\Omega_{M}^0)\, e^{3\int_0^z
d\ln(1+z')[1+w(z')]}\,, \label{friedeq}
\end{equation}
or
\begin{equation}
\frac{H^2(z)}{H_0^2} = \Omega_{M}^0 (1+z)^3 + \frac{\delta H^2}{H_0^2}\,,
\label{igno}
\end{equation}
where $H_0$ is the present value for the Hubble parameter,
$\Omega_M^0$ is the present normalized matter density and $\delta
H^2/H_0^2$ depends on the phenomenological model \cite{lindmnras}.
The EOS $w(z)$ for the dark energy can be written as
\begin{equation}
w(z)\equiv -1+{1\over3}{d\ln\delta H^2/H_0^2\over d\ln(1+z)}\,.\label{effw}
\end{equation}

The linear growth of a density fluctuation, $D= \delta\rho/\rho$,
depends on the EOS. We define the growth factor $G\equiv D/a$, where
$a\equiv 1/(1+z)$, the cosmological scale factor, and $G$ is
normalized to unity at $z\sim 1100$, the recombination epoch. In
terms of $G$, we have
\begin{equation}
G^{\prime\prime}(a)+
\left[\frac{7}{2}-\frac{3}{2}\frac{w(a)}{1+X(a)}
\right]\frac{G^{\prime}(a)}{a}+
\frac{3}{2}\frac{1-w(a)}{1+X(a)}\frac{G(a)}{a^2}=0\,,
\label{ggrowth}
\end{equation}
where $X(a)$ is defined as
\[
X(a)=\frac{\Omega_M^0 \,a^{-3}}{{\delta H^2}/{H^2_0}}\,.
\]

The linear mass power spectrum is proportional to $D^2$ and we define the deviation
from the standard $\Lambda$CDM model by
%%%%%%%%%%%%%%%%%%%%%%%%%%%%%%%%
\begin{equation}
F=\left\vert \frac{D^2 -D_{L}^{\,2}}{D^2_{L} }\right\vert\,\Big|_{z=0}\,,
\label{factor}
\end{equation}
%%%%%%%%%%%%%%%%%%%%%%%%%%%%%%%%
where $D_{ L}^2$ is the density fluctuation in the standard $\Lambda$CDM model.
Since $D^2$ derived from GD data differs from $D_{ L}^2$ derived from the CMB
anisotropies by no more than 10 per cent, the maximum value of $F$ is $F_{\rm
max}=0.1$.

We apply the above description to the following models used to explain the recent
acceleration of the universe, suggested in the literature: a phenomenological vacuum
energy decay model; two models in which gravity is modified, BWM and VMM; and the
SUGRA model.

%%%%%%%%%%%%%%%%%%%%%%% / * \ %%%%%%%%%%%%%%%%%%%%%%%%
%%%%%%%%%%%%%%%%%%%%%% /*%#%*\ %%%%%%%%%%%%%%%%%%%%%%%
%%%%%%%%%%%%%%%%%%%%%% \*%#%*/ %%%%%%%%%%%%%%%%%%%%%%%
%%%%%%%%%%%%%%%%%%%%%%% \ * / %%%%%%%%%%%%%%%%%%%%%%%%
%%% 3

\subsection{Vacuum energy decay into CDM}

In a previous paper, we put limits on the rate of a possible decay
of the vacuum energy (i.e., the decay of the cosmological constant)
into CDM from the observed agreement, to within $10\%$, of the
$(\delta\rho/\rho)^2$ derived from the CMB and the galaxy survey
data \citep{vdecay}.

Let us consider the  vacuum energy decay model described, in a flat universe, by a
power law dependence
%%%%%%%%%%%%%%%%%%%%%%%%%%%%%%%%
\begin{equation}
\Omega_{\Lambda}(z)= \Omega_{\,\Lambda}^{\,0}\,(1+z)^{\,n}\,,
\label{mainL}
\end{equation}
%%%%%%%%%%%%%%%%%%%%%%%%%%%%%%%%
where $\Omega_{\Lambda}^0=1-\Omega_M^{\,0}$. From conservation of
energy and Eq.(\ref{mainL}), we have
%%%%%%%%%%%%%%%%%%%%%%%%%%%%%%%%
\begin{equation}
\Omega_{Mv}(z)=\Omega_{M}^0 (1+z)^3-\frac{n\, \Omega_{\Lambda}^0\,}{3-n}
\left[\,(1+z)^3-(1+z)^n\right]\,, \,\label{rhopeeb}
\end{equation}
%%%%%%%%%%%%%%%%%%%%%%%%%%%%%%%%
where $\Omega_{Mv}(z=0)=\Omega^0_{M}$. Eq.(\ref{rhopeeb}) modifies
the Friedmann equation [Eq.(\ref{igno})] by a factor
%%%%%%%%%%%%%%%%%%%%%%%%%%%%%%%%
\begin{equation}
\frac{\delta H^2}{H_0^2}=\Omega_{\Lambda}^0\left[
\frac{3}{3-n}\,a^{-\,n}-\frac{n}{3-n}\,a^{-\,3} \right] \label{delta1}\,.
\end{equation}

Comparing the vacuum energy decay model with the $\Lambda$CDM model, the deviation
$F_{{\rm decay}}$ is shown in Table \ref{tabvdec}, assuming $\Omega_M^0=0.28$, the
observed value. It is to be noted that $F\lesssim 0.1$ occurs only for $n < 0.02$.

%%%%%%%%%%%%%%%%%%%%%%%%%%%%%%%%TABLE2
\begin{center}
\begin{table}[h]
\begin{tabular}{|c|c|c|c|c|c|c|c|c|}
\hline $n$ & 0.01&0.02&0.03& 0.04 &0.05& 0.06& 0.07& 0.08\\
\hline $F_{{\rm decay}}$ & 0.06 &0.12 &0.18 &0.25 &0.33 &0.41 &0.5&
0.59\\%
\hline
\end{tabular}
\caption{The deviation $F_{{\rm decay}}$ of the vacuum energy decay model from the
$\Lambda$CDM model as a function of the power $n$, assuming $\Omega_M^0=0.28$.}
\label{tabvdec}
\end{table}
\end{center}
%%%%%%%%%%%%%%%%%%%%%%%%%%%%%%%%

\begin{figure}[h]
\includegraphics[scale=0.8]{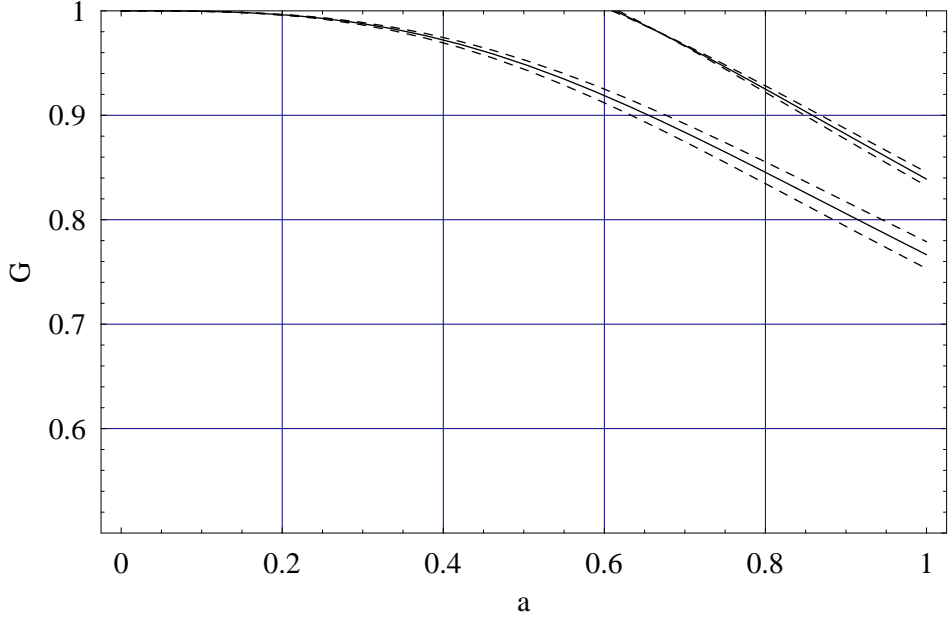}
\caption{The growth of density fluctuations,
$G=\left[\left({\delta\rho}/{\rho}\right)/a\right]$, for the $\Lambda$-decay model
with $n=0.03$ (top curve) and for the $\Lambda$CDM model (bottom curve). The dashed
lines show the deviation for the matter density $\Omega_M^0 =0.28\pm 0.02$. }
\label{figvdec}
\end{figure}

\begin{figure}[h]
\includegraphics[scale=0.8]{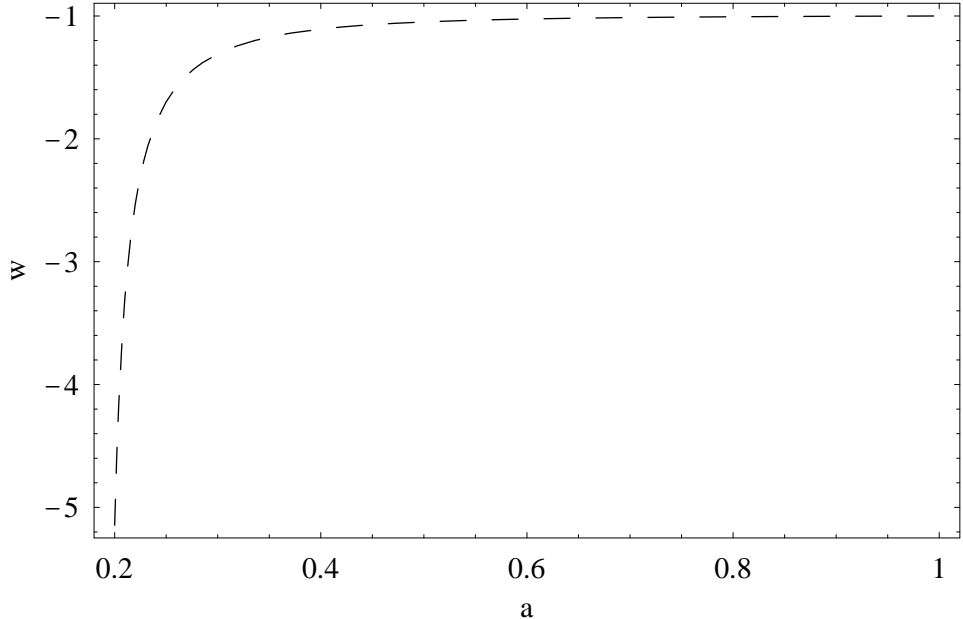}
\caption{The effective EOS, $w$, for the $\Lambda$-decay model with
$n=0.02$. } \label{weffvdec}
\end{figure}

We evaluated Eq.(\ref{ggrowth}) numerically and show $G$ as a function of $a$ for
the vacuum energy decay model in Fig.~\ref{figvdec}. A value of $n=0.03$ for the
vacuum energy decay model and $\Omega_M^0=0.28\pm 0.02$ are shown. Larger
$(\delta\rho/\rho)^2$ are predicted by the vacuum energy decay model than by the
$\Lambda$CDM model. $G$ increases with $n$ for all values of $n<3$.
%%%%%%%%%%%%%%%%%%%%%%%%%%%%%%%%%%%%%%%%%%%%%%%%FIG1
%%%%%%%%%%%%%%%%%%%%%%%%%%%%%%%%%%%%%%%%%%%%%%%%
Allowing $\Omega_M^0$ to vary for $n\geqslant  0.02$, it is not possible to obtain
an agreement with the $\Lambda$CDM within  $10\%$. The effective EOS as a function
of $a$ [Eq.(\ref{effw})] for the $\Lambda$-decay model with $n=0.02$ is shown in
Fig.~\ref{weffvdec}.
%%%%%%%%%%%%%%%%%%%%%%%%%%%%%%%%%%%%%%%%%%%%%%%%FIG2
%%%%%%%%%%%%%%%%%%%%%%%%%%%%%%%%%%%%%%%%%%%%%%%%

\subsection{Brane-world and vacuum metamorphosis models}

In the BWM \cite{bwm}, gravity is modified by adding a five-dimensional
Einstein-Hilbert action that dominates at distances which are larger than the
crossover length $r_c$ that defines an effective energy density
$\Omega_{bw}=(1-\Omega_M^0)^2/4 =1/(4H_0^2r_c^2)$ for a flat universe. The factor
$\delta H^2/H_0^2$ in Eq. (\ref{igno}) then becomes
%%%%%%%%%%%%%%%%%%%%%%%%%%%%%%%%
\begin{equation}
\delta H^2/H_0^2 = 2\Omega_{bw}+2\sqrt{\Omega_{bw}}\sqrt{\Omega_M^0(1+z)^3+
\Omega_{bw}}\,. \end{equation}
%%%%%%%%%%%%%%%%%%%%%%%%%%%%%%%%
In the VMM \cite{vmm}, the vacuum contributions are due to a quantized massive
scalar field, which is coupled to gravity. For $z<z_j$, the $\delta H^2/H_0^2$ in
Eq. (\ref{igno}) is
%%%%%%%%%%%%%%%%%%%%%%%%%%%%%%%%
\begin{equation}
\delta H^2/H_0^2 = (1-m^2/12)(1+z)^4+m^2/12-\Omega_M^0(1+z)^3\,,
\end{equation}
%%%%%%%%%%%%%%%%%%%%%%%%%%%%%%%%
where $z_j=[m^2/(3 \Omega_M^0)]^{1/3}-1$ and $m^2=3
\Omega_M^0[(4/m^2)-(1/3)]^{-3/4}$. Both the BWM and the VMM can be
described by the EOS,
%%%%%%%%%%%%%%%%%%%%%%%%%%%%%%%%
\begin{equation}
w(a)=w_0+w_a(1-a)\,, \label{lindeos}
\end{equation}
%%%%%%%%%%%%%%%%%%%%%%%%%%%%%%%%
with $(w_0,w_a)=(-0.78,0.32)$ and $(w_0,w_a)=(-1,-3)$, respectively
\cite{lindastro}.

The growth of the density fluctuation
$G=\left[\left({\delta\rho}/{\rho}\right)/a\right]$ as a function of
$a$ for the BWM, the VMM, and the $\Lambda$CDM model is shown in
Fig.~3. We note that $F$ is greater than the maximum allowed value,
$0.1$, for the BWM and the VMM.
%%%%%%%%%%%%%%%%%%%%%%%%%%%%%%%%%%%%%%%%%%%%%%%%FIG3
\begin{figure}[h]
\includegraphics[scale=0.8]{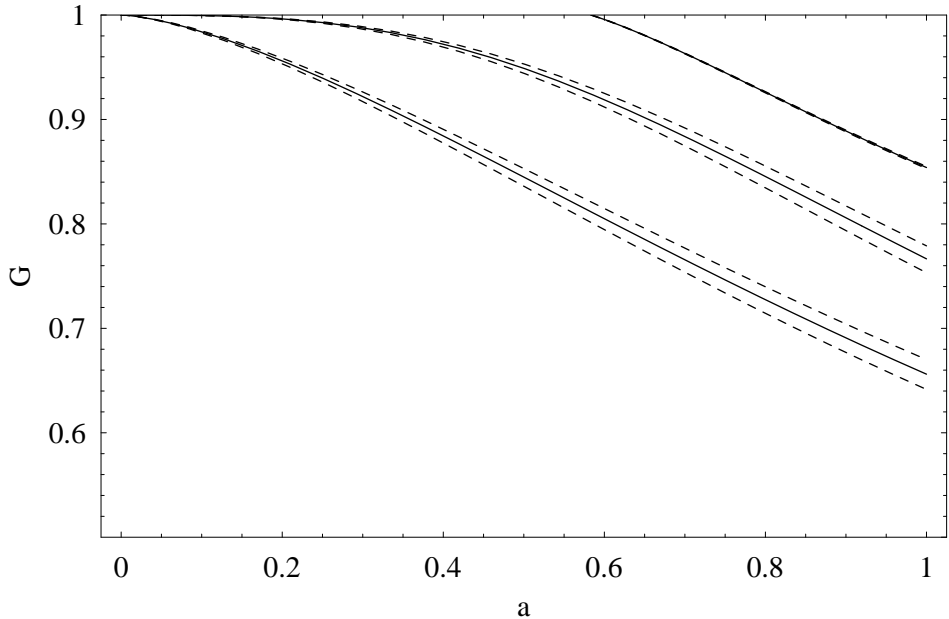}
\caption{The growth of the density fluctuation
$G=\left[\left({\delta\rho}/{\rho}\right)/a\right]$ for the VMM (top
curve), $\Lambda$CDM model (middle curve), and the BWM (bottom
curve). The dashed lines show the deviation for the matter density
$\Omega_M^0 =0.28\pm 0.02$. } \label{bwmvmm}
\end{figure}
%%%%%%%%%%%%%%%%%%%%%%%%%%%%%%%%%%%%%%%%%%%%%%%%

%%%%%%%%%%%%%%%%%%%%%%%%%%%%%%%%TABLE2
\begin{center}
\begin{table}[h]
\begin{tabular}{|c|c|c|c|c|c|}
\hline $\Omega_M^0$&$F_{\rm{BWM}}$&$H_0r_c$&$F_{\rm{VMM}}$&$m^2$&$z_j$\\
\hline
    0.26    &0.27    &1.4     &0.29    &11  &1.4\\
    0.28    &0.27    &1.4     &0.24    &11  &1.4\\
    0.3     &0.26    &1.4     &0.20    &11  &1.3\\
    0.32    &0.25    &1.5     &0.16    &11  &1.2\\
    0.34    &0.25    &1.5     &0.13    &11  &1.2\\
    0.36    &0.24    &1.6     &0.095   &10  &1.1\\
    0.72    &0.11    &3.6     &0.24    &8   &0.5\\%
\hline
\end{tabular}
\caption{The deviation $F$ for the BWM and the VMM, respectively, as a function of
the matter density $\Omega_M^0$. The values $H_0 r_0$, $m^2$ and $z_j$ are defined
in $\S$ II-B.} \label{tabbwmvmm}
\end{table}
\end{center}
%%%%%%%%%%%%%%%%%%%%%%%%%%%%%%%%
It can be seen that an agreement within  $10\%$ between the VMM and
the $\Lambda$CDM is possible only for the matter density
$\Omega_M^0\approx 0.36$. For the BWM, an agreement with the
$\Lambda$CDM within  $10\%$ is possible only if the matter density
$\Omega_M^0\approx 0.72$. Both of these values for $\Omega_M^0$ are
greater than the observational estimate $\Omega_M^0=0.28\pm 0.02$.

%%%%%%%%%%%%%%%%%%%%%%% / * \ %%%%%%%%%%%%%%%%%%%%%%%%
%%%%%%%%%%%%%%%%%%%%%% /*%#%*\ %%%%%%%%%%%%%%%%%%%%%%%
%%%%%%%%%%%%%%%%%%%%%% \*%#%*/ %%%%%%%%%%%%%%%%%%%%%%%
%%%%%%%%%%%%%%%%%%%%%%% \ * / %%%%%%%%%%%%%%%%%%%%%%%%
%%% 2
\subsection{Dark energy models described by a parametrized EOS}

We now discuss the simplest parametrization of the EOS, that has
been widely used for dark energy models, since it is well-behaved at
high redshifts (unlike $w(z)=w_0+w_1 z$ which diverges at high $z$).
This parametrization [Eq.(\ref{lindeos})], was introduced by Linder
\cite{lindprl}. The best fit parameters $w_0$ and $w_a$ that are
consistent with the Gold SNIa dataset were found to be in the
intervals $-1.91\leqslant w_0\leqslant -1.25$ and $1.53\leqslant
w_a\leqslant 5.05$ \cite{comp}. Assuming $F=0.10\pm 0.02$  and
$\Omega_M^0=0.28\pm 0.02$, we further restrain the best fit values
of $w_a$ and $w_0$. These values, whose ranges are $-1.91\le w_0\le
-1.72$ and $1.53\le w_a\le 2.9$, are shown in Table~\ref{tablinear}.
%%%%%%%%%%%%%%%%%%%%%%%%%%%%%%%%TABLE3
\begin{center}
\begin{table}[h]
\begin{tabular}{|c|c|}
\hline $w_a$    &$w_0$\\
\hline
    1.53    &$-1.72\pm 0.02$\\
    1.63    &$-1.77\pm 0.02$\\
    1.73    &$-1.82\pm 0.02$\\
    1.83    &$-1.86\pm 0.02$\\
    1.93    &-1.82-0.01\\
    2.03    &-1.89-0.01\\
    2.9 &-1.86-0.02\\%
\hline
\end{tabular}
\caption{Our best fit values of $w_0$ and $w_a$, for the EOS
$w(a)=w_0+w_a(1-a)$, for Gold SNIa dataset \cite{comp} with the
deviation $F=0.10\pm 0.02$ and the matter density
$\Omega_M^0=0.28\pm 0.02$.} \label{tablinear}
\end{table}
\end{center}
%%%%%%%%%%%%%%%%%%%%%%%%%%%%%%%%

%%%%%%%%%%%%%%%%%%%%%%% / * \ %%%%%%%%%%%%%%%%%%%%%%%%
%%%%%%%%%%%%%%%%%%%%%% /*%#%*\ %%%%%%%%%%%%%%%%%%%%%%%
%%%%%%%%%%%%%%%%%%%%%% \*%#%*/ %%%%%%%%%%%%%%%%%%%%%%%
%%%%%%%%%%%%%%%%%%%%%%% \ * / %%%%%%%%%%%%%%%%%%%%%%%%
%%% 3
\subsection{Supergravity model}

The SUGRA model \cite{sugra} is an attractive model to possibly
explain the acceleration of the universe. This model can be
described by the EOS of $\S$ II-C with $w_0=-0.82$ and $w_a=0.58$
\cite{lindmnras}. This equation of state is in agreement with
observations for the low redshift SNIa dataset \cite{sugra} and GD
data \cite{solevi}.

%%%%%%%%%%%%%%%%%%%%%%%%%%%%%%%%%%%%%%%%%%%%%%%%FIG4
\begin{figure}[h]
\includegraphics[scale=0.8]{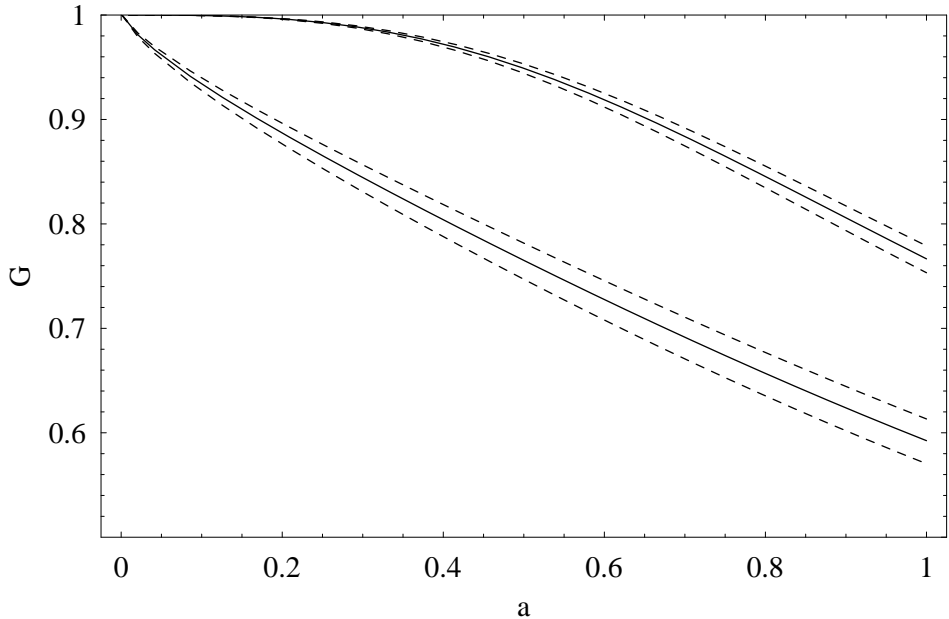}
\caption{The growth of the density fluctuation
$G=\left[\left({\delta\rho}/{\rho}\right)/a\right]$, for the
$\Lambda$CDM model (top curve) and the SUGRA model (bottom curve)
with the matter density $\Omega_M^0 =0.28\pm 0.02$ (dashed lines). }
\label{sugrafig}
\end{figure}
%%%%%%%%%%%%%%%%%%%%%%%%%%%%%%%%%%%%%%%%%%%%%%%%

Fig.~\ref{sugrafig} shows that the growth of $\delta\rho/\rho$ is
smaller for the SUGRA model than for the $\Lambda$CDM model. The $F$
for the SUGRA model is $F_{\rm SUGRA}\approx
0.38_{\,+0.02}^{\,+0.04}$ for $\Omega_M^0=0.30_{-0.02}^{-0.04}$,
which is appreciably greater than the maximum allowed value $F_{\rm
max}=0.1$.

%%%%%%%%%%%%%%%%%%%%%%% / * \ %%%%%%%%%%%%%%%%%%%%%%%%
%%%%%%%%%%%%%%%%%%%%%% /*%#%*\ %%%%%%%%%%%%%%%%%%%%%%%
%%%%%%%%%%%%%%%%%%%%%% \*%#%*/ %%%%%%%%%%%%%%%%%%%%%%%
%%%%%%%%%%%%%%%%%%%%%%% \ * / %%%%%%%%%%%%%%%%%%%%%%%%
%%% 2
\section{Conclusions}

We calculate the value of $F$ numerically for well-known dark energy
models from the growth equation for $\delta\rho/\rho$. From
observations, the maximum value of $F$ is $F_{\rm max}=0.1$. A
$\Lambda$-decay into CDM model, described by a power law dependence
(studied in our previous paper \cite{vdecay}), was first considered.
It was found that the factor $F$ increases as the exponent $n$
increases. The maximum possible value of $n$ was found to be
($n<0.02$).

The BWM and VMM were then analyzed. We showed that these models as dark energy
candidates can be discarded, assuming that $F_{\rm max}=0.1$, $\Omega_M^0=0.28\pm
0.02$ and that the universe is flat.

We combined the constraints from the Gold SNIa data \cite{comp} and the condition
that $F_{\rm max}=0.1$ to restrict the values of the parameters of the linear EOS
for dark energy, $w(a)=w_0+w_a(1-a)$. It was found that the best fit values of $w_0$
and $w_a$ are $-1.86<w_0<-1.72$ with $1.53<w_a<2.9$. For $z\sim 0.5 - 1$, where the
supernova data is sensitive, $w\sim -1$ for this parametrized EOS .

Finally, we also analyzed the SUGRA model for the above parametrized EOS with
$w_0=-0.82$ and $w_a=0.58$. $F$ was found to be very large: $F_{\rm SUGRA}\approx
0.38_{\,+0.02}^{\,+0.04}$ for $\Omega_M^0=0.30_{-0.02}^{-0.04}$, which is
appreciably greater than the maximum value $F_{\rm max}=0.1$.

\vskip 6mm
%%%%%%%%%%%%%%%%%%%%%%% / * \ %%%%%%%%%%%%%%%%%%%%%%%%
%%%%%%%%%%%%%%%%%%%%%% /*%#%*\ %%%%%%%%%%%%%%%%%%%%%%%
%%%%%%%%%%%%%%%%%%%%%% \*%#%*/ %%%%%%%%%%%%%%%%%%%%%%%
%%%%%%%%%%%%%%%%%%%%%%% \ * / %%%%%%%%%%%%%%%%%%%%%%%%
%%% 5
\noindent {\bf Acknowledgments.} R.O. thanks the Brazilian agencies FAPESP (grant
00/06770-2) and CNPq (grant 300414/82-0) for partial support. A.M.P thanks FAPESP
for financial support (grants 03/04516-0 and 00/06770-2).

%%%%%%%%%%%%%%%%%%%%%%% / * \ %%%%%%%%%%%%%%%%%%%%%%%%
%%%%%%%%%%%%%%%%%%%%%% /*%#%*\ %%%%%%%%%%%%%%%%%%%%%%%
%%%%%%%%%%%%%%%%%%%%%% \*%#%*/ %%%%%%%%%%%%%%%%%%%%%%%
%%%%%%%%%%%%%%%%%%%%%%% \ * / %%%%%%%%%%%%%%%%%%%%%%%%
%%% 6

%%%%%%%%%%%%%%%%%%%%%%%%%%%%%%%%%%%%%%%%%%%%%%%%%%%%%%%%%%%%%%%%%%%%%%%%%%%

\end{document}